\documentclass[%
preprint,
amsmath,amssymb,
aps,
]
{revtex4-1}
\usepackage{graphicx}
\usepackage{braket}
\usepackage{bm}
\usepackage{amsmath}
\usepackage{color}
\begin{document}
\title{Description of isospin mixing by a generator coordinate method}
\author{M. Kimura}
\email{masaaki@nucl.sci.hokudai.ac.jp}
\affiliation{Department of Physics, Hokkaido University, Sapporo 060-0810, Japan} 
\affiliation{Nuclear Reaction Data Centre, Hokkaido University, Sapporo 060-0810, Japan} 
\affiliation{Research Center for Nuclear Physics (RCNP), Osaka University, Ibaraki 567-0047, Japan}
\author{Y. Suzuki}
\affiliation{Department of Physics, Hokkaido University, 060-0810 Sapporo, Japan}
\author{T. Baba}
\affiliation{Kitami Institute of Technology, 090-8507 Kitami, Japan}
\author{Y. Taniguchi}
\affiliation{Department of Information Engineering, National Institute of Technology (KOSEN), Kagawa
College, 769-1192 Mitoyo, Japan} 
\affiliation{Research Center for Nuclear Physics (RCNP), Osaka University, Ibaraki 567-0047, Japan}  
\date{\today}

\begin{abstract}
 \noindent{\bf Background}: The isospin mixing is an interesting feature of atomic nuclei. It plays
 a crucial role in the astrophysical nuclear reactions.  However, it is not straightforward for
 variational nuclear structure models to describe it.

 \noindent{\bf Purpose}: We propose a tractable method to describe the isospin mixing within a
 framework of generator coordinate method, and demonstrate its usability. 

 \noindent {\bf Method}: We generate the basis wave functions by applying the
 Fermi transition operator to the wave functions of isobars. The superposition of these basis wave
 functions and  variationally obtained wave functions quantitatively describes the isospin mixing. 

 \noindent {\bf Results}: Using $^{14}{\rm N}$ as an example, we demonstrate that our method
 reasonably describes both $T=0$ and 1 states and their mixing. Energy spectrum and $E1$
 transition strengths are compared with the experimental data to confirm isospin mixing.

 \noindent{\bf Conclusion}: The proposed method is effective enough to describe isospin mixing and
 is useful, for example, when we discuss $\alpha$ capture reactions of $N=Z$ nuclei. 
\end{abstract}

\maketitle

\section{introduction}
The isospin symmetry is a fundamental symmetry of strong interaction and nuclear force. Because of
this symmetry, isobars share a group of states having the same value of the total isospin, which are
called isobaric analog states~\cite{Warburton1969,Bohr1969}. For example, an $N=Z$ nucleus
$^{14}{\rm N}$ has  the $T=1$ states such as the $0^+_1$ state at 2.31 MeV and the $1^-_2$ state at
8.06 MeV which are the isobaric analog states corresponding to the ground and first excited states
of $^{14}{\rm C}$ and  $^{14}{\rm O}$. 

In reality,  isospin is an approximate symmetry of atomic nuclei due to the symmetry
breaking terms of nuclear force and Coulomb interaction. Consequently, the mixing of the states with
different isospin (isospin mixing) occurs especially in the excited states. Above mentioned $1^-_2$
state of $^{14}{\rm N}$ is a well known example of the isospin mixing for which the admixture of the
$T=0$ and 1 states are experimentally confirmed~\cite{Barker1966,Renan1972}. 

An interesting side effect of isospin mixing is that on the selection rule of the $E1$
transitions~\cite{Warburton1969}. If there is no isospin mixing, the $E1$ transitions between two $T=0$ states
are  forbidden because $E1$ transition operator is purely isovector at the first order of the long 
wave-length approximation. However, with isospin mixing, the transition occurs due to the
small contamination of the $T=1$ components. It is notable that this allowed $E1$ transition
sometimes plays a crucial role in astrophysical reactions~\cite{Rolfs1988,RevModPhys.74.1015}.  The
radiative $\alpha$ capture reactions of  $N=Z$ nuclei such as 
$^{12}{\rm C}(\alpha,\gamma)^{16}{\rm O}$ reaction~\cite{Deboer2017} are famous examples for this.
reactions. The isospin mixing increases the reaction rate, and can affect the 
evolution of stars and the abundance of the elements.  

Thus, the isospin mixing in $N=Z$ nuclei is an interesting issue relevant to astronomical nuclear
reactions. However, the description of the isobaric analog states and isospin mixing are not
straight forward for the variational models such as Hartree-Fock models. Since the $T=0$ states are
usually more deeply bound than the $T=1$ states, the energy variation yields only the $T=0$ states
and the $T=1$ states are hardly obtained. To overcome this problem, several methods and
prescriptions have been suggested~\cite{Satua2001,Sato2013,Kanada-Enyo2015,Baczyk2018}. For example,
the isospin projection before the variation~\cite{Satua2010,Morita2016} is a solid approach to this
problem but computationally demanding. Therefore, the development of a simpler but accurate method
is desirable.   

In this paper, we propose a tractable method to describe isobaric analog states and isospin mixing
within a framework of generator coordinate method. We generate the basis wave functions by applying
the Fermi transition operator to the wave functions of isobars. Using $^{14}{\rm N}$ as an example,
it will be shown that the superposition of thus-obtained wave functions and  variationally obtained
wave functions quantitatively describes both $T=0$ and 1 states and the isospin mixing.

This paper is organized as follows. In the next section, we introduce a method to describe isobaric
analog states and isospin mixing. In the section~\ref{sec:result}, we present the numerical results
for the $T=0$ and 1 states and their mixing in $^{14}{\rm N}$ . Final section summarizes this work.  

\section{Theoretical framework}
\subsection{Hamiltonian and variational wave function}
We use the $A$-body Hamiltonian given as,
\begin{align}
 H=-\sum_{i}^{A} \frac{\hbar^2\nabla_i^2}{2m_N}-t_{\rm cm}
 +\sum_{i<j}^{A} v_{NN}(ij)
 +\sum_{i<j}^{Z} v_{\rm C}(ij),
\end{align}
where the Gogny D1S density functional~\cite{Berger1991} is used as an effective nucleon-nucleon 
interaction ($v_{NN}$), and proton-neutron mass difference is ignored. In other words, we
only consider the Coulomb interaction as the source of isospin symmetry breaking. This
simplification may be validated in the case of $^{14}{\rm N}$ which we will discuss later, because
the Coulomb interaction should dominate over other symmetry breaking terms. 

The variational wave function is a parity-projected Slater determinant,
\begin{equation}
\Phi^{\pi}=\hat{P}^\pi \mathcal{A} \{\varphi_{1}\varphi_{2}\dots\varphi_{A}\}, \quad \pi=\pm
\end{equation}
where $\hat{P}^\pi$ is the parity projection operator. The single-particle wave packet $\varphi_{i}$ 
is represented by a deformed Gaussian~\cite{Kimura2004a},
\begin{align}
 \varphi_i(\bm{r}) &= \prod_{\sigma=x,y,z}
 e^{-\nu_\sigma \left (r_{\sigma}-Z_{i\sigma} \right)^2}\chi_{i} \eta_{i},\\
\chi_{i}&=a_{i}\chi_\uparrow+b_{i}\chi_\downarrow, \quad
 \eta_{i}= \set{\mathrm{proton\ or\ neutron}}.
\end{align}
The variational parameters are the width $(\nu_x, \nu_y, \nu_z)$ and the centroids $\bm Z_i$ of
Gaussian wave packets, and spin direction $a_i$ and $b_i$. They are determined by the energy
variation with the constraint on the matter quadrupole deformation parameter $\beta$. We denote the
wave function obtained by the energy variation as $\Phi^\pi(\beta)$ which has the minimum energy for
given value of the parameter $\beta$. 
Note that the energy variation tends to yield the wave functions with minimum isospin as they are
energetically favored. In short, the energy variation produces wave functions with $T=0$ for
$^{14}{\rm N}$ and those with $T=1$ for $^{14}{\rm C}$. 

\subsection{Basis wave functions for isobaric analog states}
As explained above, it is not straightforward to obtain the wave functions of isobaric analog
states, e.g., the $T=1$ states of $^{14}{\rm N}$, by the energy variation. Here, we propose a simple
method to generate the basis wave functions for describing the isobaric analog states. Let us explain it
by using the $T=1$ states of $^{14}{\rm N}$ as examples.  Suppose that we have obtained the
wave function of $^{14}{\rm C}$ by the energy variation,
\begin{align}
 \Phi^\pi(\beta,{}^{14}{\rm C}(T=1))
 =P^\pi\mathcal{A}\set{\varphi_1\cdots\varphi_6\varphi_7\cdots\varphi_{14}},
\end{align}
where $\varphi_1\cdots\varphi_6$ and $\varphi_7\cdots\varphi_{14}$ are the proton and neutron
single-particle wave packets, respectively. Note that this wave function is mostly composed of the
$T=1$ component (minimum isospin for $^{14}{\rm C}$). Then, we simply apply the Fermi transition
operator of $\beta^-$ decay to produce the wave function of $^{14}{\rm N}$,
\begin{align}
 \Phi^\pi(\beta,{}^{14}{\rm N}(T=1)) =  T^+\Phi^\pi(\beta,{}^{14}{\rm C}(T=1))\nonumber\\
 =\sum_{i=7}^{14}P^\pi\mathcal{A}
 \set{\varphi_1\cdots\varphi_6\varphi_7\cdots\varphi_i\cdots\varphi_{14}},\label{eq:iaswf1}
\end{align}
where $i$th neutron wave packet ($i=7,...,14$) is turned into proton. Since $T^+$ commutes with
$T^2$, this wave function approximates isobaric analog state ($T=1$ states) of $^{14}{\rm N}$. We
propose to use each term of Eq.~(\ref{eq:iaswf1}) as the basis wave 
function for generator coordinate method. Thus, we generate eight wave functions for $^{14}{\rm N}$
from a single $^{14}{\rm C}$ wave function, which are denoted as,
\begin{align}
 \Phi^\pi_i(\beta,{}^{14}{\rm N}(T=1)) =  t^+_i\Phi^\pi(\beta,{}^{14}{\rm C}(T=1))\nonumber\\
 =P^\pi\mathcal{A} \set{\varphi_1\cdots\varphi_6\varphi_7\cdots\varphi_i\cdots\varphi_{14}}.
 \label{eq:iaswf2}
\end{align}

\subsection{Generator coordinate method}
Once the basis wave functions are prepared, we perform the angular momentum projection and GCM
calculation. The basis wave functions are projected to the eigenstates of the angular momentum and
superposed, 
\begin{align}
 \Psi^{J\pi}_{M}=&\sum_{\beta K}f_{\beta K} P^{J}_{MK}\Phi^\pi(\beta,{}^{14}{\rm N}(T=0)) \nonumber\\
 &+ \sum_{\beta K i}g_{\beta K i} P^{J}_{MK}\Phi^\pi_i(\beta,{}^{14}{\rm N}(T=1)),
\end{align}
where $P^{J}_{MK}$ is angular momentum projection operator. Note that 
$\Phi^\pi(\beta,{}^{14}{\rm N}(T=0))$ are  obtained by the energy variation and mainly consist of
$T=0$ components whereas $\Phi^\pi_i(\beta,{}^{14}{\rm N}(T=1))$  are generated by
Eq.~(\ref{eq:iaswf2}) and predominated by $T=1$ components.  The eigen-energies and the coefficients
of superposition $f_{\beta K}$ and $g_{\beta Ki}$ are determined by solving the Hill-Wheeler
equation~\cite{Hill1953}.

\section{results and discussion}\label{sec:result}
\begin{figure}[hbt]
\centering
\includegraphics[width=0.5\hsize]{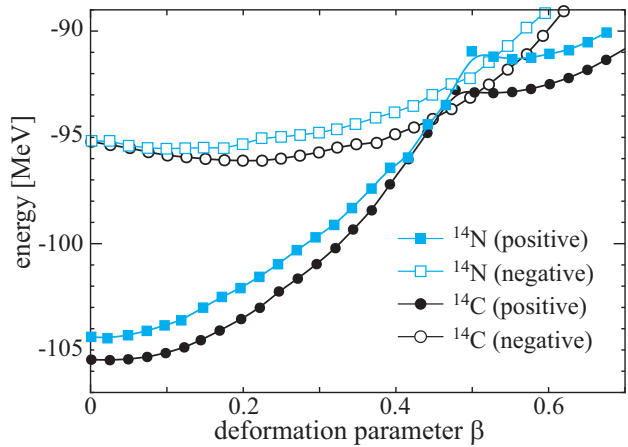}
\caption{Energy curves for $^{14}{\rm N}$ and $^{14}{\rm C}$ as functions of the quadrupole
 deformation parameter $\beta$ obtained by the energy variation after parity projection.}
\label{fig:curve}
\end{figure}

Figure~\ref{fig:curve} shows the energy curves of $^{14}{\rm N}$ and  $^{14}{\rm C}$ as functions
of the quadrupole deformation parameter $\beta$ which are obtained by the energy variation after the
parity projection.  
The positive-parity states have the spherical energy minimum with $N=8$ closed shell
for $^{14}{\rm C}$ and a single neutron hole for $^{14}{\rm N}$ which are the dominant components of
the ground states. As  quadrupole deformation grows, the energy increases rapidly and the level
crossing occurs creating energy plateau around $\beta=0.6$. In this plateau, both nuclei have
two-particle and two-hole ($2p2h$) configurations which generate the highly excited 
states~\cite{Suhara2010,Baba2016}. The negative-parity energy curves are located at much higher
energy than the positive-parity states as they involve particle-hole excitation across the $N=Z=8$
shell gap. Although it is not clear from the shape of the energy curves, the negative-parity states
have  $1p1h$ configurations in the small deformation region, and  $3p3h$ configurations in the
largely deformed region. Thus, both nuclei have similar behavior of the energy curve due to  the
similarity in the single particle configurations.  The energy difference between $^{14}{\rm N}$ and
$^{14}{\rm C}$ is due to the difference in the the isospin channel ($T=0$ for $^{14}{\rm N}$ and
$T=1$ for  $^{14}{\rm C}$) and the Coulomb interaction.    

As explained in the previous section, we apply the Fermi transition operator of $\beta^-$ decay to
the wave functions of $^{14}{\rm C}$ (circles in Fig.~\ref{fig:curve}) to yield the wave
functions of $^{14}{\rm N}$ with $T=1$. The wave functions thus-generated are superposed with the
wave functions of $^{14}{\rm N}$ with $T=0$ (squares in Fig.~\ref{fig:curve}) to
perform the GCM calculations.

\begin{figure*}[hbt]
\centering
\includegraphics[width=0.9\hsize]{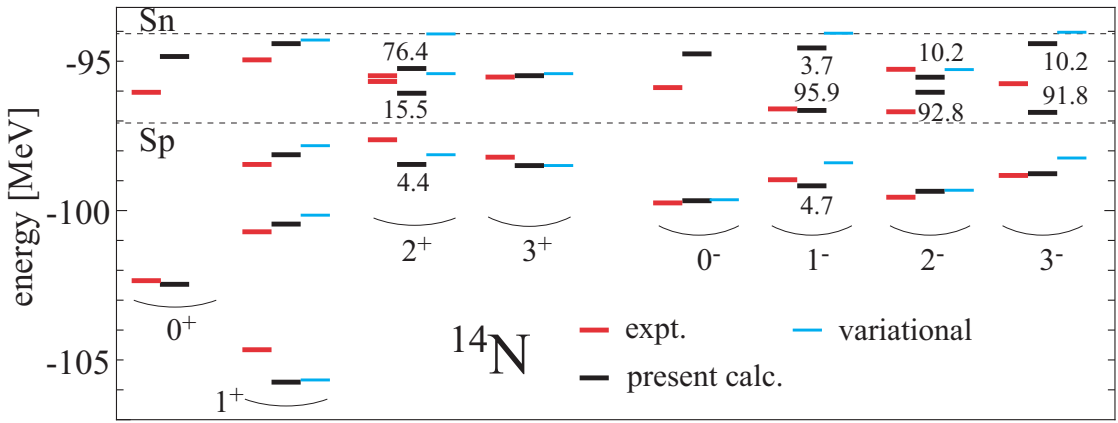}
\caption{Calculated and observed level scheme of $^{14}{\rm N}$ up to $J^\pi\leq 3^\pm$. Black lines
 show the result  of the present model whereas blue lines show the results calculated by using only
 the variationally  obtained basis wave functions. Numbers in the figure shows the amount of the
 $T=1$ component in percentage for the states with sizable isospin mixing.}
\label{fig:level}
\end{figure*}

The spectra of $^{14}{\rm N}$ obtained by the GCM calculations are shown in Figure~\ref{fig:level}.
By using only the variationally obtained wave functions of $^{14}{\rm N}$ (without the $T=1$ wave
functions generated from $^{14}{\rm C}$), the GCM calculation (blue lines in Fig.~\ref{fig:level})
fails to reproduce several excited states such as the  $0^+_{1,2}, 0^-_2$ and $1^-_2$ states, all of
which are  $T=1$ states. Thus, the variational calculations energetically favor the $T=0$ states
and leaves out the $T=1$ states. On the contrary, by adding the basis wave functions generated by
applying the Fermi transition operator to the $^{14}{\rm C}$ wave functions,  the present model
plausibly describes both $T=0$ and 1 states (black lines). Note that the all the observed states up
to $E_x<10$ MeV are reasonably reproduced by our simple method, although the calculation slightly
overestimate the binding energy of the ground state. To elucidate the accuracy of the present
calculations, Table~\ref{tab:em} shows the electromagnetic properties of the $1^+_{1,2}$ $(T=0)$ and
$0^+_1$ $(T=1)$ states. It is encouraging that the present calculation reproduces not only
the electromagnetic moments of the individual states but also the transition probabilities between
$T=0$ and 1 states~\cite{Ajzenberg-Selove1991}. Thus, our proposed method is simple but accurately
describes both of the $T=0$ and 1 states with a small computational cost.  
 
\begin{table}[hbt]
 \caption{The calculated and observed \cite{Ajzenberg-Selove1991} properties of the low-lying $1^+$
 and $0^+$  states. Excitation energy, magnetic dipole and electric quadrupole moments are given in
 the unit of  MeV, $\mu_N$ and $e\rm fm^2$, respectively. The $M1$ and $E2$ reduced transition
 probabilities are  given in the Weisskopf unit.} \label{tab:em}     
 \begin{ruledtabular}
  \begin{tabular}{ccccccc}
   &\multicolumn{2}{c}{$E_x$}&\multicolumn{2}{c}{$\mu$}&\multicolumn{2}{c}{$Q$}\\
   \hline
   $J^\pi$ & calc. & exp. & calc. & exp. & calc. & exp. \\
   \hline
   $1^+_1(T=0)$ & 0.0 & 0.0 & 0.39 & 0.40 & 1.7 & 1.9 \\
   $1^+_2(T=0)$ & 3.2 & 2.3 & 0.45 & 0.39 & 0.8 & $-$ \\
   $0^+_1(T=1)$ & 4.9 & 3.9 & $-$  & $-$  & $-$  & $-$ \\
  \end{tabular}
  \vspace{0.1cm}
  \begin{tabular}{ccccc}
   &\multicolumn{2}{c}{$B(M1)$}&\multicolumn{2}{c}{$B(E2)$}\\
   \hline
   $J_i^\pi\rightarrow J_f^\pi$ & calc. & exp. & calc. & exp. \\
   \hline
   $1^+_1\rightarrow 0^+_1$ & 0.023              & 0.026 & $-$ & $-$ \\
   $1^+_2\rightarrow 0^+_1$ & 1.2                & 1.0   & $-$ & $-$ \\
   $1^+_1\rightarrow 0^+_2$ & 3.3$\times10^{-4}$ & 1.4$\times10^{-4}$&
   2.2 & 2.7
  \end{tabular}
 \end{ruledtabular}
\end{table}

Now, we discuss the isospin mixing described by our model. The present calculation yielded several
states with sizable isospin mixing, for which we showed the mixing ratio (amount of the $T=1$
component in percentage) in Fig.~\ref{fig:level}. It is notable that the states close to the  
$^{13}{\rm C}+p$ or $^{13}{\rm N}+n$  threshold energies have strong isospin mixing, which is
related to the origin of the isospin mixing as explained below. 

\begin{table}[hbt]
 \caption{The electric and magnetic dipole transition strengths for the $1^-_1$ and $1^-_2$ 
 states~\cite{Renan1972,Ajzenberg-Selove1991} given in Weisskopf unit.} \label{tab:BEM}
\begin{ruledtabular}
 \begin{tabular}{cccc}
  $J_i^\pi\rightarrow J_f^\pi$ & $|\Delta T|$&$B(E1)_{\rm exp}$ & $B(E1)_{\rm calc}$ \\
  \hline
  $1^-_1 \rightarrow 0^+_1$ & 1 & $\geq 2.4\times 10^{-3}$     & $3.2\times 10^{-3}$ \\
  $1^-_1 \rightarrow 1^+_1$ & 0 & $\geq 2.3\times 10^{-4}$     & $1.8\times 10^{-4}$ \\
  $1^-_1 \rightarrow 1^+_2$ & 0 & -                            & $1.5\times 10^{-4}$ \\
  $1^-_2 \rightarrow 0^+_1$ & 0 & $(2.3\pm 0.7)\times 10^{-3}$ & $1.6\times 10^{-3}$ \\
  $1^-_2 \rightarrow 1^+_1$ & 1 & $(4.8\pm 1.2)\times 10^{-2}$ & $3.6\times 10^{-2}$ \\
  $1^-_2 \rightarrow 1^+_2$ & 1 & $(5.7\pm 1.5)\times 10^{-2}$ & $3.4\times 10^{-2}$ \\
  \hline
  $J_i^\pi\rightarrow J_f^\pi$ & $|\Delta T|$ &$B(M1)_{\rm exp}$ & $B(M1)_{\rm calc}$ \\
  \hline
  $1^-_1 \rightarrow 0^-_1$ & 0 & -    & 0.01 \\
  $1^-_1 \rightarrow 2^-_1$ & 0 & -    & 0.06 \\
  $1^-_2 \rightarrow 0^-_1$ & 1 & 1.6  & 1.6\\
  $1^-_2 \rightarrow 1^-_1$ & 1 & 0.06 & 0.17\\
  $1^-_2 \rightarrow 2^-_1$ & 1 & 0.36 & 0.36\\
 \end{tabular}
 \end{ruledtabular}
\end{table}

Among these states, the $1^-_1$ (5.69 MeV, $T=0$) and $1^-_2$ (8.06 MeV, $T=1$) states are
well-known example of the isospin mixing for which the mixing ratio was evaluated 
experimentally~\cite{Renan1972}. In the following, we discuss how these states are described by our model
and anatomy their structure. Table~\ref{tab:BEM} lists the electric and magnetic dipole
transition probabilities of these $1^-$ states. Firstly, note that the experimental values of the
electric and magnetic dipole transition strengths of the $1^-$ states are reproduced
accurately. This indicates that our model 
precisely describes the wave functions of the $1^-$ states. Secondly, we should focus on the
intensities of the $E1$ transition strengths. Since $^{14}{\rm N}$ is a self-conjugate nucleus,
isospin selection rule allows only $|\Delta T|=1$ transitions and forbids the $|\Delta T|=0$
transitions~\cite{Warburton1969}. In fact, the $|\Delta T|=0$ transitions are suppressed by an order
of magnitude compared to the $|\Delta T|=1$ transitions for both observed and calculated values. At
the same time, non-zero values for the $|\Delta T|=0$ transitions indicate the isospin mixing in the
$1^-_1$ and $1^-_2$ states. In Ref.~\cite{Renan1972}, the mixing ratio was estimated from the $E1$
transition probabilities. They assumed that the $1^+_{1,2}$ and $0^+_1$ have no isospin mixing and
the $1^-_{1,2}$ states are admixture of two components as,
\begin{align}
 \Psi(1^-_1) &=\alpha \Phi(1) + \beta \Phi(0),\\
 \Psi(1^-_2) &=\beta \Phi(1) - \alpha \Phi(0),
\end{align}
where $\Phi(0)$ and $\Phi(1)$ denote the $T=0$ and 1 wave functions. Then, the ratio of the allowed
and forbidden transitions gives an estimate of the mixing ratio as,
\begin{align}
 \frac{B(E1)_{\rm forbidden}}{B(E1)_{\rm
 allowed}}=\frac{\alpha^2}{\beta^2}=\frac{\alpha^2}{1-\alpha^2}.
\end{align}
They adopted the observed $1^-_2\rightarrow 1^+_1$ and $1^-_2\rightarrow 0^+_1$ transition strengths
and obtained an estimate of $\alpha^2_{\rm exp}=0.046$ whereas our calculated transition strengths
yield $\alpha^2_{\rm calc}=0.042$, both of which are close to the mixing ratio directly calculated
from our $1^-$ wave functions ($\alpha^2=0.047$ for $1^-_1$ and 0.041 for $1^-_2$). Other
combinations of the transition strengths also suggest similar values, for example, $1^-_1\rightarrow
1^+_1$ and $1^-_1\rightarrow 0^+_1$ transitions give $\alpha^2_{\rm exp}=0.09$ and $\alpha^2_{\rm
calc}=0.053$.

In order to understand the origin of isospin mixing, we investigate the spectroscopic factors and
the overlap functions. We calculate the overlap between the wave function of $^{14}{\rm N}$ 
with the spin-parity ${J'}^{\pi'}$ and that of $^{13}{\rm N}$ with $J^\pi$~\cite{Kimura2017},
\begin{align}
 \varphi(\bm r)=
 \sqrt{13} \braket{\Psi^{{J\pi}}_{M'-m}(^{13}{\rm N})|\Psi^{J'\pi'}_{M'}(^{14}{\rm N})}.
\end{align}
The overlap function is given as the multipole decomposition of $\varphi(\bm r)$,
\begin{align}
 \varphi_{jl}(r) = \int d\hat r [Y_{l}(\hat r)\times\chi_{1/2}]^\dagger_{jm}\varphi(\bm r),
\end{align}
which is the radial wave function of a valence neutron in the $J^{\pi}\otimes \nu(l_j)$ channel.  
The spectroscopic factor is  the norm of $\varphi_{jl}(r)$, 
\begin{align}
 S(J^{\pi}\otimes \nu(l_j)) = \int r^2 dr|\varphi_{jl}(r)|^2.
\end{align}

\begin{table}[hbt]
 \caption{The spectroscopic factors of the $1^+_1$, $0^+_1$ and $1^-_{1,2}$ states in the  
 $J^{\pi}\otimes \nu(l_j)$ and $J^{\pi}\otimes \pi(l_j)$ channels, where ${J}^{\pi}$ denotes
 the spin-parity of $^{14}{\rm N}$ and $^{14}{\rm C}$ whereas $l_j$ denotes the orbital and total
 angular momenta of a valence neutron or proton.} \label{tab:sfac}
\begin{ruledtabular}
 \begin{tabular}{ccccc}
  &$\frac{1}{2}^-\otimes \nu(p_{1/2})$& 
   $\frac{1}{2}^-\otimes \pi(p_{1/2})$& 
   $\frac{5}{2}^-\otimes \nu(p_{3/2})$& 
   $\frac{5}{2}^-\otimes \pi(p_{3/2})$\\
  $1^+_1$ & 0.87 & 0.87 & 1.28 & 1.29 \\
  \hline
  &$\frac{1}{2}^-\otimes \nu(p_{1/2})$& 
   $\frac{1}{2}^-\otimes \pi(p_{1/2})$& 
   $\frac{3}{2}^-\otimes \nu(p_{3/2})$& 
   $\frac{3}{2}^-\otimes \pi(p_{3/2})$\\
  $0^+_1$ & 0.90 & 0.90 & 0.99 & 1.04 \\
  \hline
  &$\frac{1}{2}^-\otimes \nu(s_{1/2})$& 
   $\frac{1}{2}^-\otimes \pi(s_{1/2})$& 
   $\frac{1}{2}^+\otimes \nu(p_{1/2})$& 
   $\frac{1}{2}^+\otimes \pi(p_{1/2})$\\
  $1^-_1$ & 0.23 & 0.43 & 0.47 & 0.25 \\
  $1^-_2$ & 0.38 & 0.23 & 0.28 & 0.42 \\
 \end{tabular}
 \end{ruledtabular}
\end{table}

\begin{figure}[hbt]
\centering
\includegraphics[width=0.5\hsize]{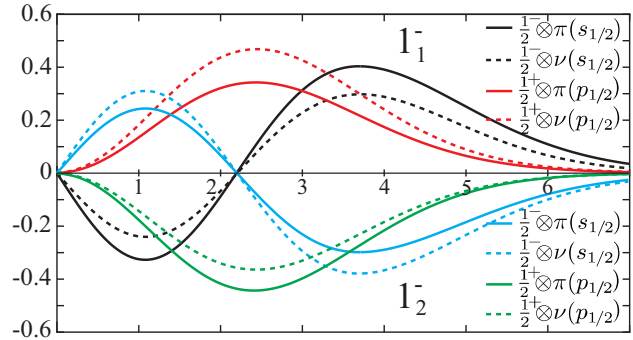}
\caption{The overlap functions of the $1^+_1$, $0^+_1$ and $1^-_{1,2}$ states in the  
 $J^{\pi}\otimes \nu(l_j)$ and $J^{\pi}\otimes \pi(l_j)$ channels. The phases of the overlap
 functions are arbitrary chosen for the presentation.} \label{fig:ofunc}
\end{figure}

The spectroscopic factors in the $J^{\pi}\otimes \pi(l_j)$ channels (the overlap between 
$^{14}{\rm N}$ and $^{13}{\rm C}$) are also calculated in the same manner.
The calculated spectroscopic factors and the overlap functions of the $1^-$ states are given in
Table~\ref{tab:sfac} and Fig.~\ref{fig:ofunc}, respectively. For comparison, we also present
spectroscopic factors for the $1^+_{1} (T=0)$ and $0^+_{1} (T=1)$ states which have no isospin mixing. 
If we assume that both $^{13}{\rm C}$ and $^{13}{\rm N}$ are the eigenstates of isospin with
$T=1/2$, which is a reasonable assumption indeed, the spectroscopic factors in the 
$J^\pi\otimes \nu(l_j)$ and $J^\pi\otimes \pi(l_j)$ channels should be equal for the pure $T=0$
or 1 states. In fact, we see the equality holds for the low-lying $1^+_1$ and $0^+_1$
states. However, we found that the spectroscopic factors for the $1^-_1$ and $1^-_2$ states show
significant asymmetry between two channels because of the isospin
mixing. The $1^-_1$ state has larger contribution from the $1/2^-\otimes \pi(s_{1/2})$ and
the $1/2^+\otimes \nu(p_{1/2})$ channels compared to the $1/2^-\otimes \nu(s_{1/2})$ and
the $1/2^+\otimes \pi(p_{1/2})$ channels, whereas the $1^-_2$ state shows the
opposite trend.   

The origin of the asymmetry is understood as follows. Notice that $^{13}{\rm N}(1/2^-)$ and
$^{13}{\rm N}(1/2^+)$ are approximated as the $\nu(p_{1/2})$ and $\nu(s_{1/2})$ states on top of the
$^{12}{\rm C}$ ground state as an inert core. The same also applies to $^{13}{\rm C}(1/2^{\pm})$. 
So, we see that both $1/2^-\otimes \nu(s_{1/2})$ and $1/2^+\otimes \pi(s_{1/2})$ channels are
identically represented as $\pi(s_{1/2})\nu(p_{1/2})$. In the same manner, 
$1/2^+\otimes \nu(s_{1/2})$ and $1/2^-\otimes \pi(p_{1/2})$ are
represented as $\pi(p_{1/2})\nu(s_{1/2})$. Hence, we understand that
Tab.~\ref{tab:sfac} implies that the $1^-_1$ and $1^-_2$ are approximated by the linear combinations 
of $1p1h$ excitation across the $N=Z=8$ shell gap,
\begin{align}
 \ket{1^-_1} &= a\ket{\pi(s_{1/2})\nu(p_{1/2})}-b\ket{\pi(p_{1/2})\nu(s_{1/2})}, \\
 \ket{1^-_2} &= b\ket{\pi(s_{1/2})\nu(p_{1/2})}+a\ket{\pi(p_{1/2})\nu(s_{1/2})},
\end{align}
with $a>b$, which is consistent with the assumption made in the shell model
calculations~\cite{Warburton1960,Sebe1963,Hsieh1970,Jager1971}. 
Since the ground state of $^{14}{\rm N}$ be approximated as
$\pi(p_{1/2})\nu(p_{1/2})$, this indicates that the proton excitation 
$\pi(p_{1/2}\rightarrow s_{1/2})$ is energetically favored over the neutron excitation
 $\nu(p_{1/2}\rightarrow s_{1/2})$. This owes to the Coulomb energy difference. As shown in
 Fig.~\ref{fig:ofunc}, the $\pi(s_{1/2})$ orbit is 
spatially extended than $\pi(p_{1/2})$ as it is close to the threshold energy. Consequently, 
$\pi(s_{1/2})$ has smaller Coulomb repulsion than $\pi(p_{1/2})$, and hence, the single-particle
excitation energy which costs for $\pi(p_{1/2}\rightarrow s_{1/2})$ is reduced than that for
$\nu(p_{1/2}\rightarrow s_{1/2})$. We also found that other excited states with 
isospin mixing such as $2^+_{1,2,3}$, $2^-_{2,3}$ and $3^-_{2,3}$ states always involve the
single-particle excitation to $s_{1/2}$. Therefore, we conclude the Coulomb energy shift of
proton $s_{1/2}$ orbit is a major source of the isospin mixing in the excited states close to the
proton and neutron decay thresholds. 

\section{summary}\label{sec:summary}
In this work, we proposed a tractable method to describe the isospin mixing within a framework of
generator coordinate method. By applying the Fermi transition operator to the wave functions of
isobars, we generated the wave functions of the isobaric analog states which are used as the basis
of GCM calculations. Using $^{14}{\rm N}$ as an example,  we demonstrated that our tractable method
plausibly describes both of $T=0$ and 1 states and the isospin mixing in the excited states close to
the proton and neutron thresholds. We showed that our model reasonably describes the strengths of
the allowed and forbidden $E1$ transitions that are consistent with the mixing ratio. Furthermore,
based on the spectroscopic factors and overlap functions, we deduced that the Coulomb energy shift of
$s_{1/2}$ orbit is a major source of the isospin mixing. 

\acknowledgements
This work was supported by the JSPS KAKENHI Grant No. 19K03859. Part of the numerical computation in
this work was carried out at the Yukawa Institute Computer Facility.

\bibliography{n14}

\end{document}